\documentclass[review]{elsarticle}

\usepackage{lineno,hyperref}
\usepackage{float}
\usepackage[T1]{fontenc}
\usepackage{fourier}
\usepackage[english]{babel}
\usepackage{amsmath,amsthm}
\usepackage{graphicx}
\usepackage{color}
\usepackage[justification=justified]{caption}
\usepackage{subcaption}
\usepackage{comment}
\usepackage{amsthm}
\usepackage{tikz}
\usepackage{enumitem}
\usetikzlibrary{shapes,arrows}
\usetikzlibrary{calc}
\usetikzlibrary{positioning, decorations.pathmorphing, shadows,decorations.markings}
\usepackage{pgf,tikz}
\usetikzlibrary{arrows.meta, calc, chains, quotes, positioning, shapes.geometric, fit}
\usepackage{booktabs}
\theoremstyle{definition}

\modulolinenumbers[5]

\journal{Computational Materials Science}

\begin{document}

\begin{frontmatter}

\title{Learning to Fail: Predicting Fracture Evolution in Brittle Material Models using Recurrent Graph Convolutional Neural Networks}

\author[cgu,pomona]{Max Schwarzer}
\author[cgu,pomona]{Bryce Rogan}
\author[cgu]{Yadong Ruan}
\author[cgu]{Zhengming Song}
\author[cgu,sdsu]{Diana~Y.~Lee}
\author[cgu]{Allon G.~Percus\corref{cor}}
\ead{allon.percus@cgu.edu}
\author[eeslanl]{Viet T.~Chau}
\author[broadridge]{Bryan A.~Moore}
\author[eeslanl]{Esteban~Rougier}
\author[eeslanl]{Hari S. Viswanathan}
\author[xcplanl]{Gowri Srinivasan}

\address[cgu]{Institute of Mathematical Sciences, Claremont Graduate University, Claremont, CA 91711, USA}
\address[pomona]{Mathematics Department, Pomona College, Claremont, CA 91711, USA}
\address[sdsu]{Computational Science Research Center, San Diego State University, San Diego, CA 92182, USA}
\address[eeslanl]{Earth and Environmental Sciences Division, Los Alamos National Laboratory, Los Alamos, NM 87545, USA}
\address[broadridge]{Broadridge Financial Solutions, New York, NY 10016, USA}
\address[xcplanl]{X Computational Physics Division, Los Alamos National Laboratory, Los Alamos, NM 87545, USA}

\cortext[cor]{Corresponding author}

\begin{abstract}
We propose a machine learning approach to address a key challenge in materials science: predicting how fractures propagate in brittle materials under stress, and how these materials ultimately fail.  Our methods use deep learning and train on simulation data from high-fidelity models, emulating the results of these models while avoiding the overwhelming computational demands associated with running a statistically significant sample of simulations. We employ a graph convolutional network that recognizes features of the fracturing material and a recurrent neural network that models the evolution of these features, along with a novel form of data augmentation that compensates for the modest size of our training data.  We simultaneously generate predictions for qualitatively distinct material properties. 
Results on fracture damage and length are within 3\% of their simulated values, and results on time to material failure, which is notoriously difficult to predict even with high-fidelity models, are within approximately 15\% of simulated values.  Once trained, our neural networks generate predictions within seconds, rather than the hours needed to run a single simulation.
\end{abstract}


\end{frontmatter}



\section{Introduction}

Brittle materials, such as ceramics, glasses, and some metal alloys, are important in a variety of contexts. By definition, these are materials that break with little plastic deformation, meaning that they fracture without bending or stretching to more than a minimal extent. Brittle material failure occurs under stress and can result in catastrophic damage, in turn leading to significant financial and human loss. Historical failures such as the Schenectady Liberty ship and Aloha Airlines Flight 243 demonstrate this clearly: the Schenectady broke nearly in half within only 16 days of service, due to the brittle fracture of low-grade steel components, and the Boeing 737 used for Flight 243 suffered extensive damage due to multisite fatigue cracking of the skin panels, which in turn caused in-flight explosive decompression at 24,000 feet~\cite{Fleming2013}. 

Microstructural information and the dynamic process of fracture propagation play key roles in understanding the mechanisms of brittle material failure, enabling predictions on whether, when, and how material failure might occur. Applications of fracture dynamics include hydraulic mining in shale gas systems~\cite{HymanJimenez2016}, fluid migration~\cite{Kim2014}, and aircraft safety assessment~\cite{White2006}.
Numerous methods have been developed to model and study these dynamics. A fruitful approach has been the use of hybrid finite-discrete element methods (FDEM)~\cite{Hillerborg1976}, including the Hybrid Optimization Software Suite (HOSS)~\cite{Rougier2014}.  This high-fidelity model simulates fracture and fragmentation processes or materials deformation in both 2D and 3D complex systems, providing accurate predictions of fracture growth leading to material failure.  However, running such simulations can require many thousands of hours for materials on the scale of meters, making them computationally prohibitive.

In order to reduce computation time dramatically while retaining accuracy, we propose a machine learning approach to predict dynamic fracture propagation.  Our method uses deep learning to model simultaneously a broad range of evolving features in a fracturing material.   We convert HOSS simulation results into training data for a neural network that learns the temporal dynamics of a fracture system.  Processing the data and training the neural network can both be parallelized efficiently.  While this process can take on the order of hours, subsequently using the trained network on an initial set of microcracks can yield predictions within seconds for fracture growth statistics and time to material failure.  We thereby speed up computation by many orders of magnitude.

Our approach involves transforming the HOSS FDEM mesh into a fracture graph, where vertices represent fractures and edge weights represent relationships between fractures~\cite{Vevatne2014,Valera2018}.
As features in the graph, we define fracture attributes that include physical features such as fracture length, damage, tip stress, position and orientation, as well as connectivity metrics that include coalescence, damage between fractures, and multiple forms of physical distance.  Based on a single graph and its features, our goal is to produce a full time series of graphs that predicts the evolution of these features, mirroring the full dynamics simulated by HOSS.  The fracture graph representation also results in a vast reduction of data size, from nearly a terabyte of data for the output of a set of HOSS simulations to only about 20GB for a set of graph time series.

We use deep learning to predict the evolution of our fracture graphs.  Deep learning uses large artificial neural networks for capturing complex nonlinear relationships hidden in data, and has been successful in areas such as automatic speech recognition, image recognition, and natural language processing.  Neural networks are well-suited to our problem because a single network can predict the simultaneous evolution of multiple features.  Our network architecture exploits the capabilities of a graph convolutional network (gCNs)~\cite{kipf2016semi} to recognize features from the reduced graph representation of large fracture networks, coupled with a recurrent neural network (RNN)~\cite{manessi2017dynamic} to model the evolution of those features.

We train our network on fracture graph time series from 145 HOSS simulations.  Once trained, the network takes a $t=0$ graph as input and outputs its prediction for the graph at all subsequent time steps $t=1,2,\dots,n$.  We find that these predictions closely match the actual statistics of HOSS simulations.  The predicted fracture size is, on average, within 2\% of its simulated value, and the predicted distribution of fracture sizes is consistent with the simulated distribution (at the 0.05 significance level) in 87\% of the simulations.  For the time to failure, the mean absolute error of predictions is approximately 15\% of the average.

In Section~\ref{backgroundsection}, we provide the physical and algorithmic background for our approach.  Section~\ref{physicalsystemsection} describes the physical system and the HOSS simulations representing it. Section~\ref{graphsection} gives the reduced graph representation we use for our fracture systems, and Section~\ref{methodssection} discusses our ML methods.  Section~\ref{resultssection} presents the results of these methods, and Section~\ref{conclusions} summarizes our work and discusses its implications.

\section{Background}
\label{backgroundsection}

HOSS~\cite{Rougier2014} is a fracture simulator based on the combined finite-discrete element method (FDEM).  FDEM has been used since the 1990s~\cite{Munjiza1995,Munjiza2004,Lei2014} to model fragmentation in brittle materials, notably the transition from continuum to discrete behavior that takes place when the material fails.  The FDEM formulation describes fracture and fragmentation processes explicitly based on conservation laws, avoiding unnecessary assumptions regarding the behavior of the model.

HOSS simulations have been used to study fracture initiation and propagation in shale rocks in order to enhance the efficiency of hydraulic fracturing~\cite{2014AGUFM} and to model earthquake ruptures~\cite{2017EGUGA}.  While highly accurate, HOSS faces limitations due to computational intensity.  Solid domains are discretized, with each fracture represented by tens to hundreds of finite elements.  A material on the scale of millimeters can contain $10^6$ microcracks, and require $10^7$-$10^9$  elements~\cite{djidjev_omalley_viswanathan_hyman_karra_srinivasan_2017}. Because of the explicit scheme used by HOSS, with Newton's laws providing the governing equations, correctly resolving the physics of such a material requires very small time steps.  Consequently, even simulations involving laboratory-size samples of several centimeters, with thousands of microcracks, can require thousands of hours of computation to model milliseconds of dynamics and create petabytes of data. Additionally, the sample's initial configuration of defects and microcracks is rarely known precisely.  This necessitates an uncertainty quantification framework, requiring thousands of simulations to provide statistically significant results on system behavior.  Such an approach becomes computationally intractable for materials of practical interest, which are often on the scale of meters.

Machine learning (ML) methods can help mitigate the need for overwhelming computational power in modeling fracture dynamics, by rapidly predicting the outcome of simulations. ML techniques such as support vector machines \cite{SHI2008} and artificial neural networks in various forms \cite{Wang2017} have been applied to fracture analysis, demonstrating promising performance. The use of high-fidelity fracture network simulations as training data together with novel graph representations was successfully applied to identify areas of high flow and transport~\cite{Valera2018} in static discrete fracture networks generated by the dfnWorks suite~\cite{Hyman2015}. For the problem of dynamic graph representations of fractures, Miller et al.~\cite{Miller2017} has employed an image-based approach, extracting video representations of HOSS simulations and using graph convolutional networks to learn features from these.  Finally, Moore et al.~\cite{Moore2018} and Srinivasan et al.~\cite{Srinivasan2018} have used a variety of ML methods to predict fracture coalescence and time to failure.  These studies suggest that important aspects of fracture dynamics can be learned from a modest sample of simulation training data, and then predicted through modern algorithmic techniques.

\begin{figure}[!htb]
\centering
\includegraphics[height=0.5\linewidth]{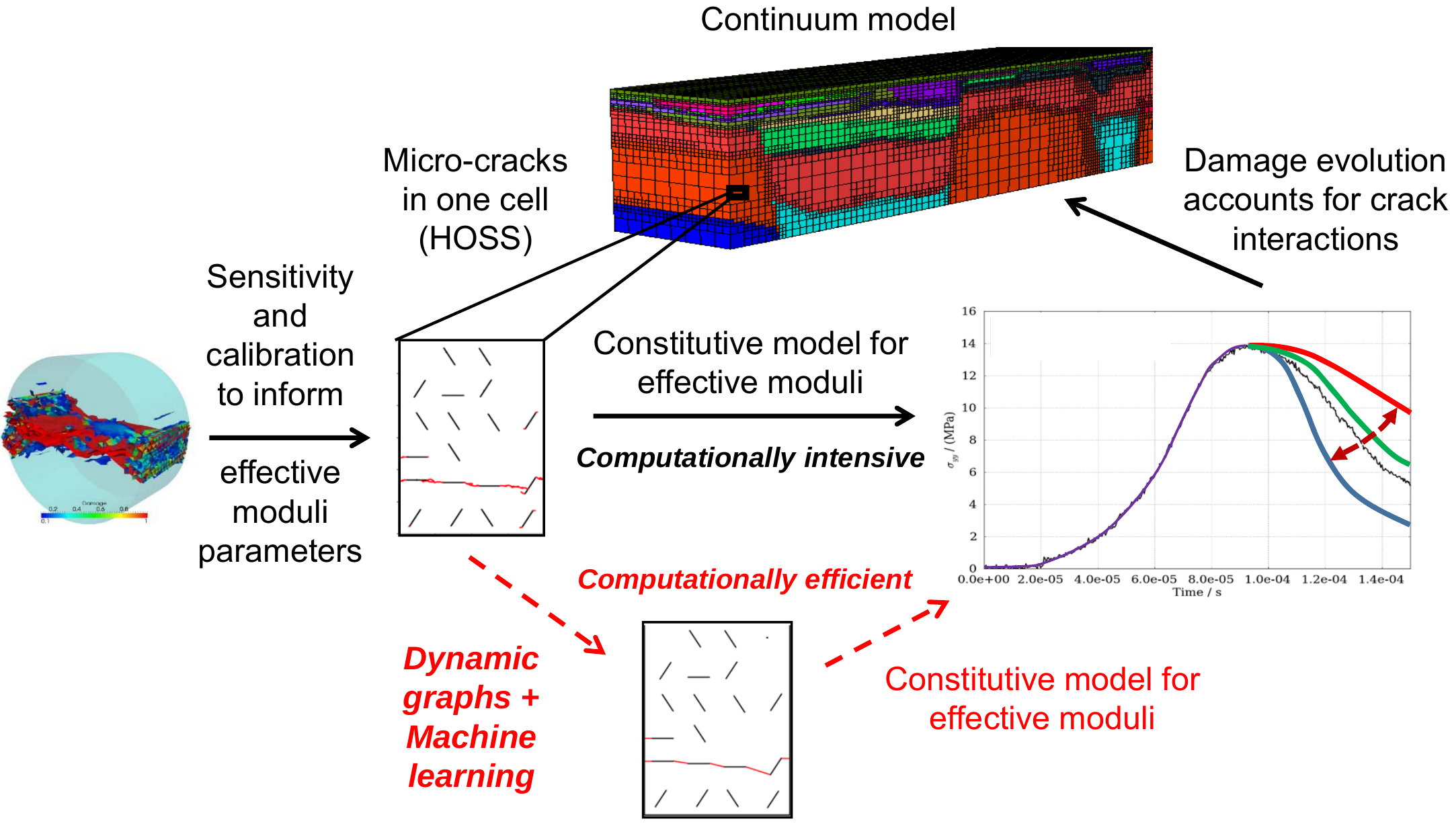}

\caption{Our workflow uses HOSS to provide training data for ML rather than for direct simulation, yielding a computationally more efficient approach to predicting fracture propagation.}
\label{fig:workflow}
\end{figure}

Figure~\ref{fig:workflow} summarizes the workflow in our approach: from a continuum model, to HOSS simulations over a statistical ensemble, to the use of ML on dynamic graphs for vastly accelerated predictions of fracture statistics and material failure.

\section{Physical System and HOSS Simulations}
\label{physicalsystemsection}

\begin{figure}[!htb]
\centering
\includegraphics[width=0.5\linewidth]{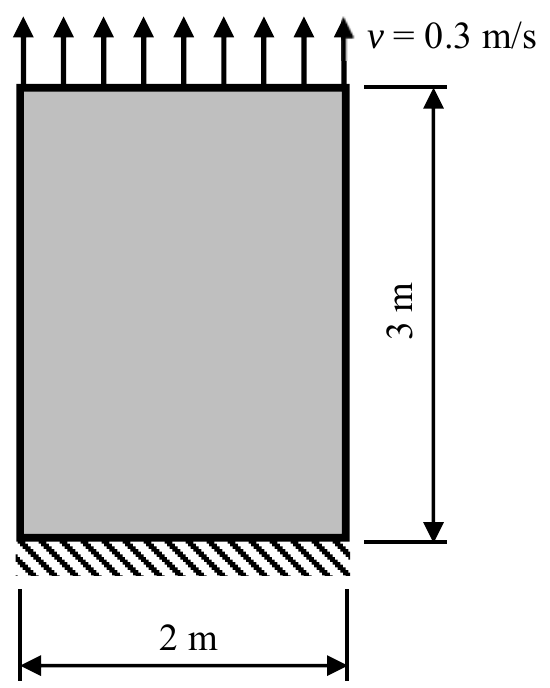}
\caption{General setup for the physical model, with boundary and loading conditions as indicated.}
\label{fig:setup}
\end{figure}

The high-fidelity HOSS simulations, used for training our algorithms, represent a system composed of a rectangular 2m $\times$ 3m 2D concrete sample loaded in uniaxial tension. In order to impose this loading condition, the bottom boundary of the sample is kept fixed while the top boundary is moved with a constant speed of 0.3m/s, as shown in figure~\ref{fig:setup}.  This results in a strain rate of 0.1 s$^{-1}$.  The material is assumed to be elastically isotropic for all cases.  The following elastic material properties were used throughout: density of 2500 kg/m$^3$, Young's modulus of 22.6 GPa, and shear modulus of 9.1 GPa.

Each specimen is seeded with 20 fractures at time $t=0$, representing preexisting microcracks or defects, each of length 30cm.
With the aim of avoiding overlaps among these initial fractures, the sample was divided into 24 cells arranged in a uniform $4\times 6$ grid, of which 20 cells were selected randomly to contain a fracture. The orientation of each fracture was chosen randomly from one of three angles: 0, 60 and 120 degrees with respect to the horizontal.  Finally, each fracture was positioned randomly in its cell, subject to the constraint of being fully contained within the cell.

\begin{figure}[!t]
\centering
\includegraphics[height=0.25\linewidth]{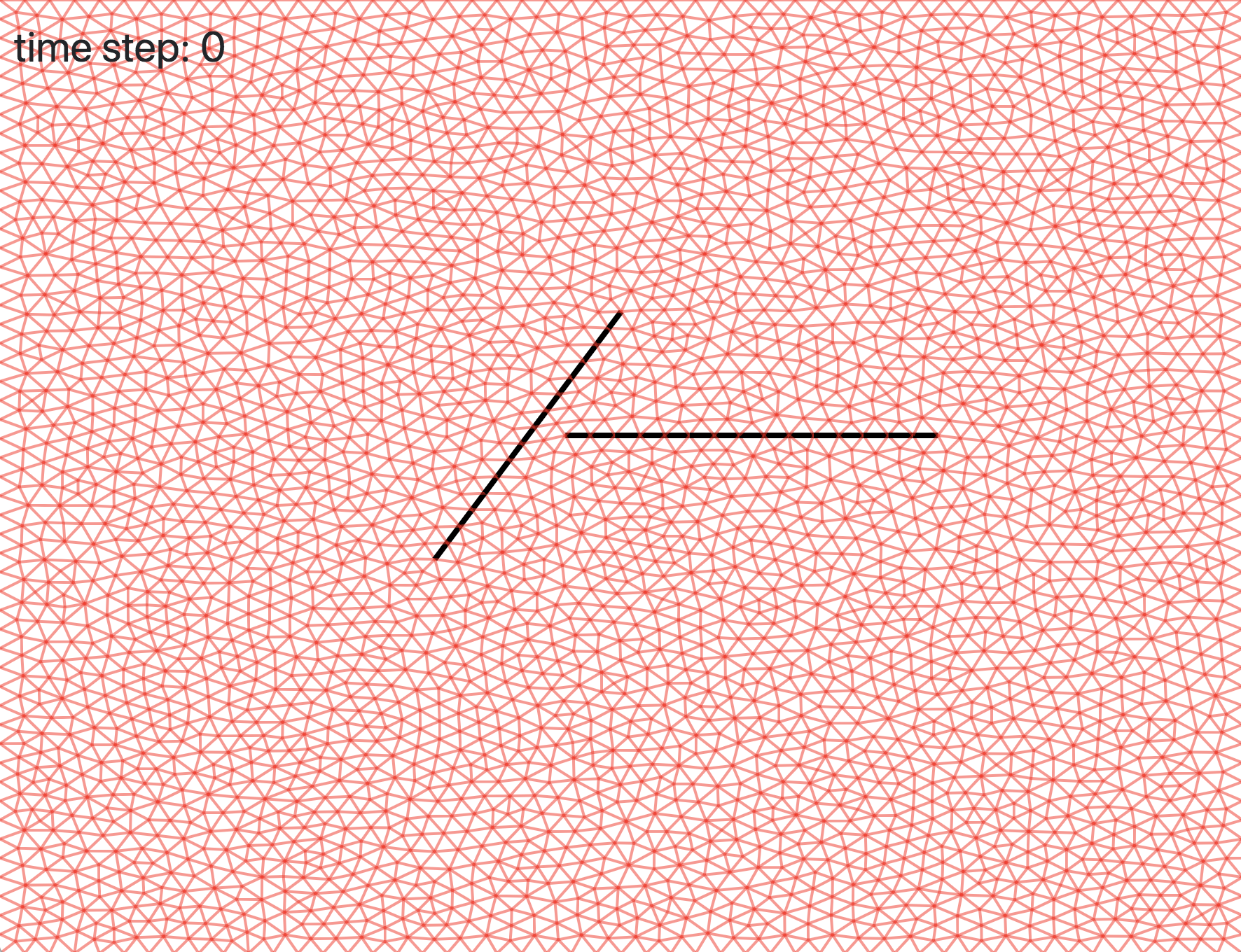}\;
\includegraphics[height=0.25\linewidth]{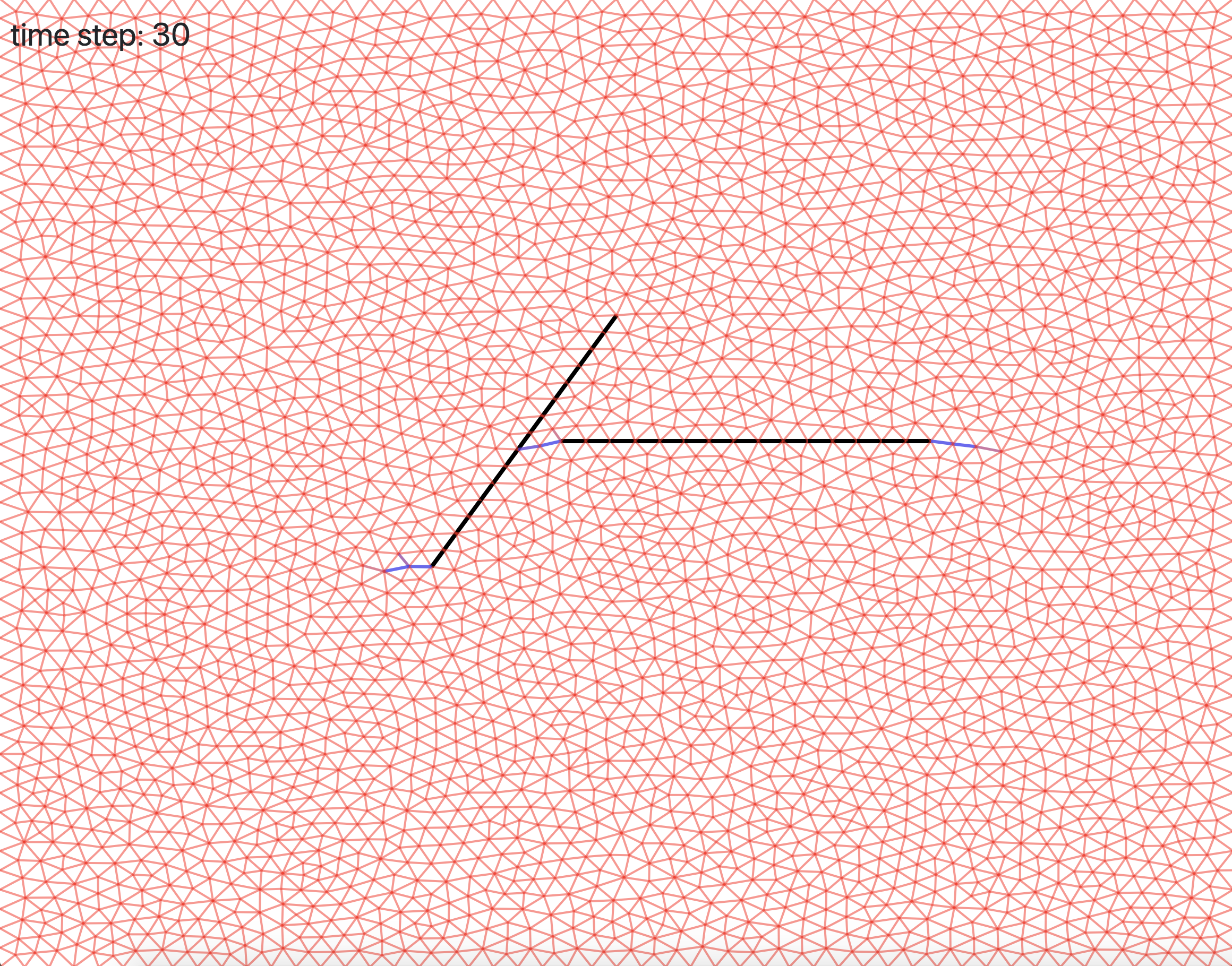}\;
\includegraphics[height=0.25\linewidth]{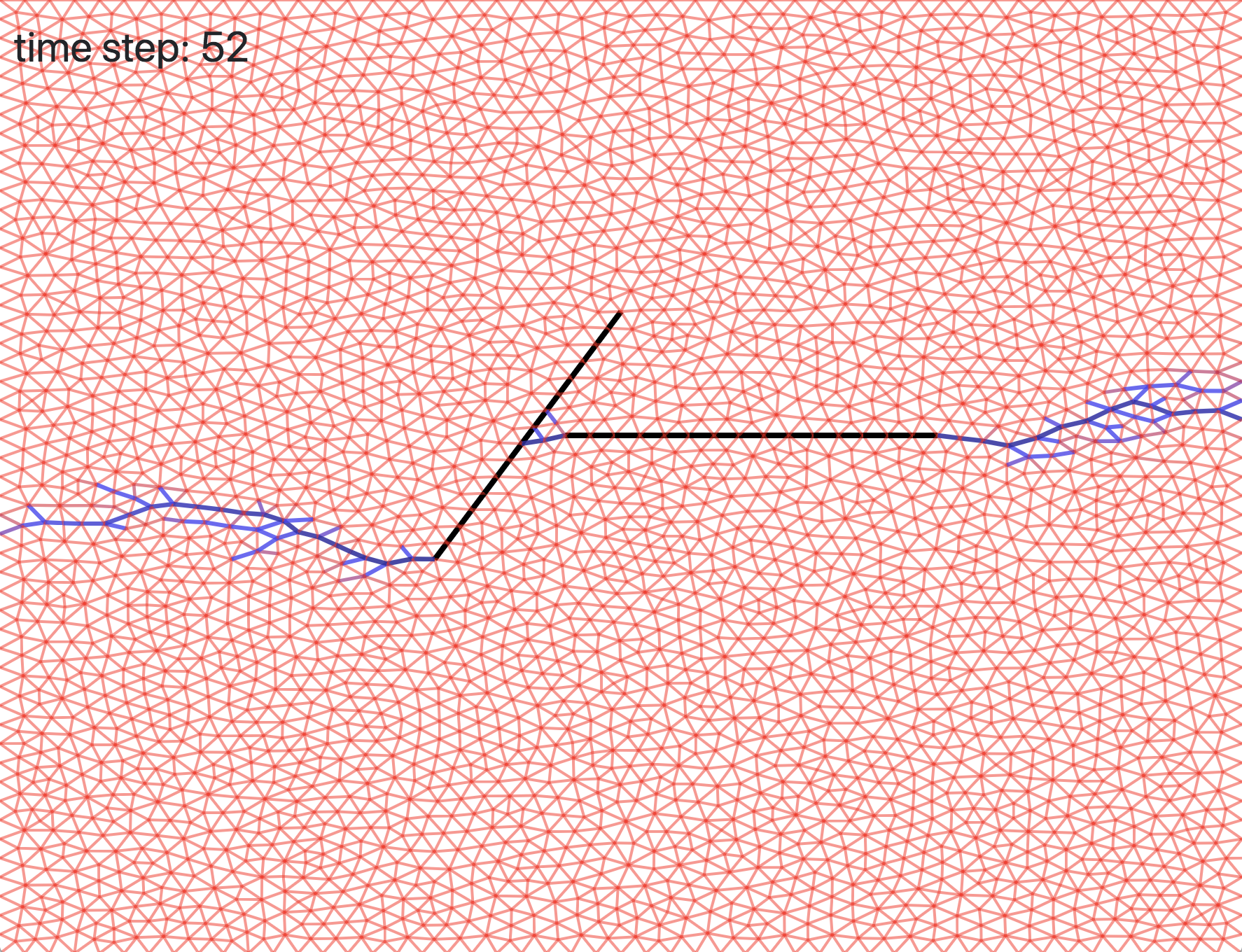}

(a) \hspace{1.375in} (b) \hspace{1.375in} (c)

\caption{HOSS triangulation of two-fracture system.  (a) Black edges represent initial fracture nuclei.  (b) Fractures expand and coalesce.  (c) Material failure occurs when fractures connect to boundary.}
\label{fig:time}
\end{figure}

HOSS simulations use finite elements to model the complex physical process of fracture propagation, with fractures moving along the edges of a triangulated mesh and coalescence points forming at the junctions (figure~\ref{fig:time} shows an example in the simpler case of only two fractures).  The system described above was discretized using approximately 158,000 constant strain triangular elements per specimen. Each vertex of a triangle can move independently in 2D, leading to nearly $10^6$ degrees of freedom.  Fractures are characterized by a damage value associated with each edge, ranging from $0$ (completely intact) to $1$ (completely broken).  When the damage exceeds a fixed threshold $\theta$, set to $\theta=0.1$, that edge is considered to be fractured.

Simulations were run for 350 time steps, representing 7 milliseconds of fracture evolution.  Each simulation took approximately 4 hours to run, using 400 processors.

\section{Graph Formulation}
\label{graphsection}

%
%
%

Our approach involves converting the finite element representation to a fracture graph, where vertices represent fractures, labels associated with each vertex represent fracture attributes (such as length) and edges represent relationships between fractures.

\subsection{Fracture Graph}
\label{fracturegraphsection}

We extract the fracture graph by identifying connected components of fractures in the finite element mesh.  In order to obtain meaningful predictions of fracture propagation from graph-based neural networks methods, we would like the total number of vertices in the graph to remain fixed as the system evolves and these fractures coalesce.  We therefore adopt the following rules:

\begin{enumerate}
\item Fracture identity is preserved through coalescence.  If two fractures $i$ and $j$ coalesce at time step $t$, they nevertheless retain their identities at time step $t+1$.  This allows each one to retain its associated features, which will be crucial for our neural network approach.  It is important to be able to determine the point of coalescence so that, as the fractures grow, one can decide which finite elements have been newly absorbed into which fractures. This may be ambiguous if fractures grow rapidly between time steps, or if three or more fractures coalesce in one time step.  In those cases, we assign the finite element to the fracture where the least additional damage is needed to connect it to its previous elements (see minimal damage distance in section~\ref{ConnectivitySection} below).

\item A finite element associated with a fracture at a given time step remains associated with that fracture for all future time steps.  We assume that fractures can only grow over time, 
\end{enumerate}

\begin{figure}
	\centering
    \hspace{-20pt}
    \includegraphics[width=0.5\linewidth]{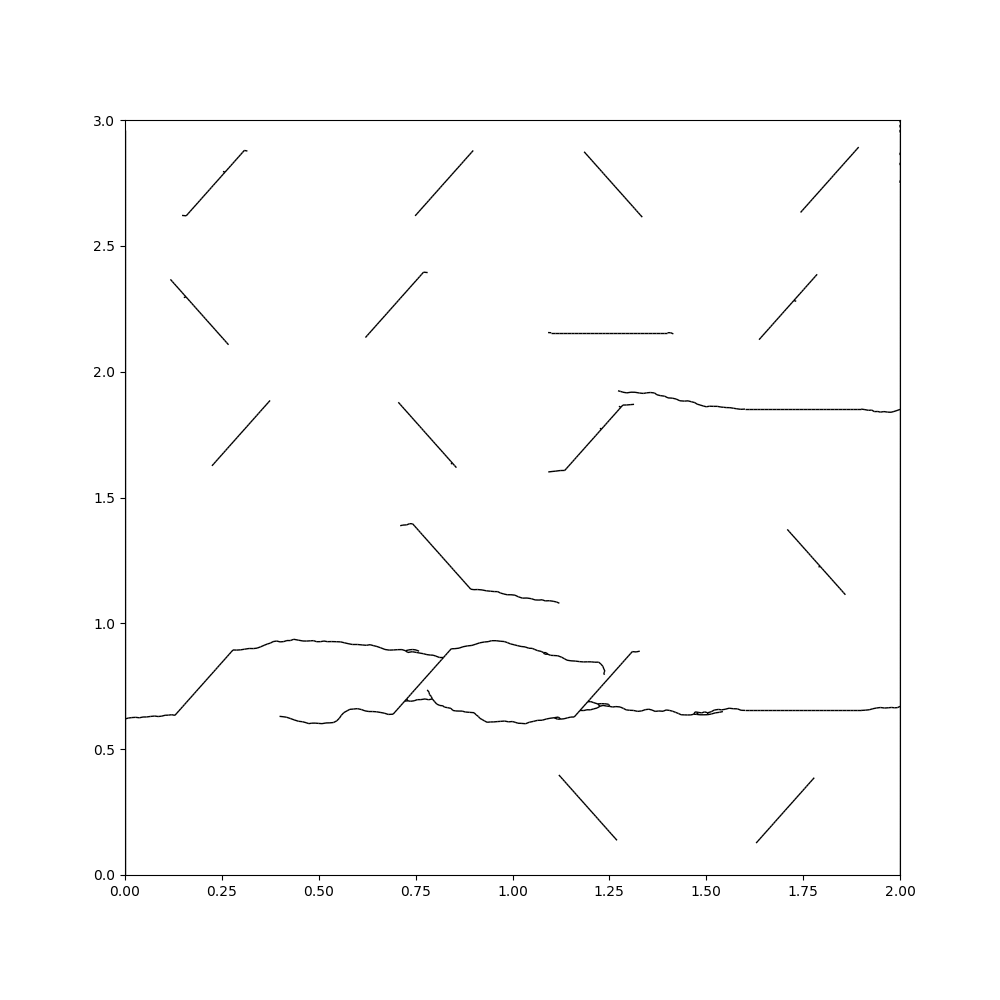}
    \raisebox{17pt}{\includegraphics[width=0.4\linewidth]{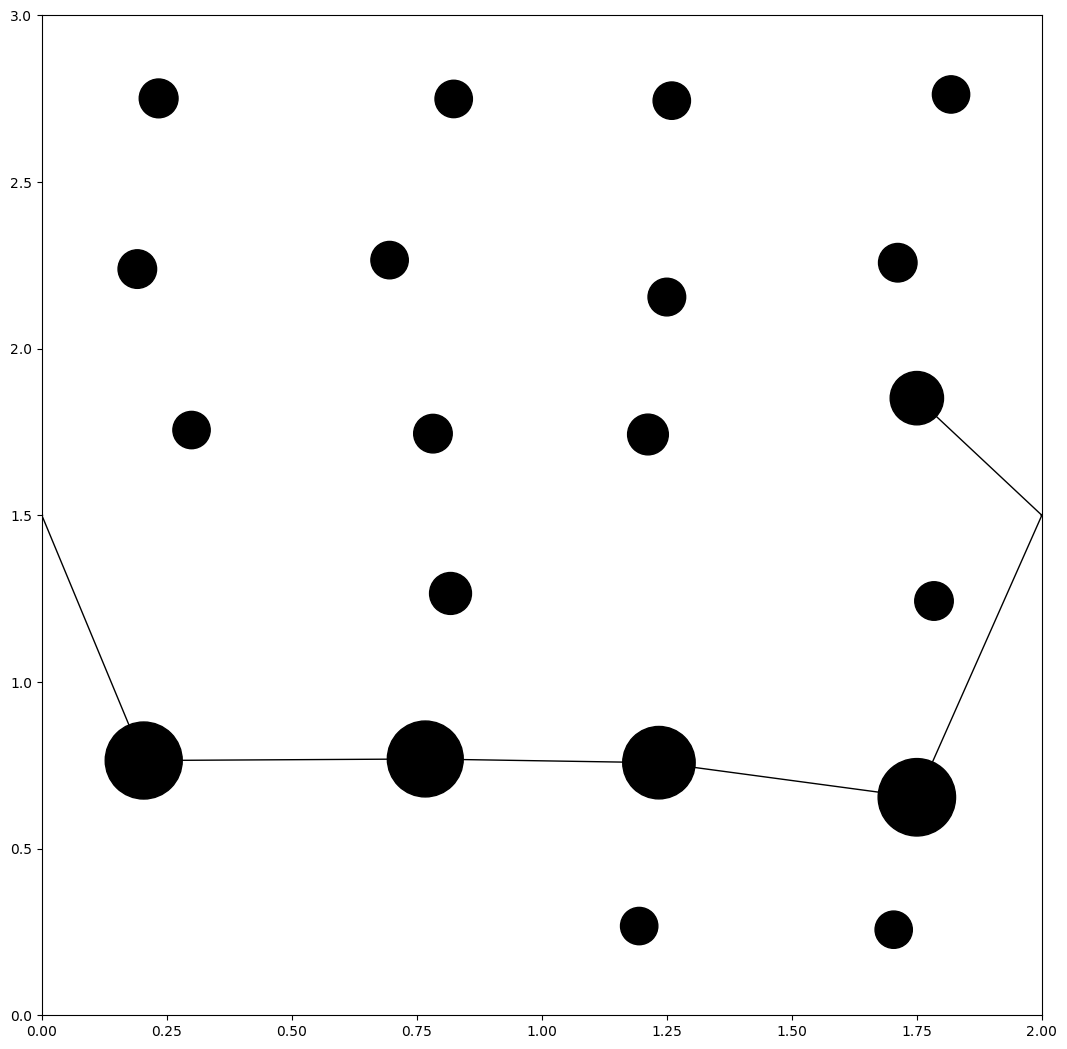}}

    \vspace{-20pt}
    
    (a) \hspace{2in} (b)

	\caption{A fracture system produced by HOSS, and the corresponding fracture graph representation.  (a) Lines represent fractures, defined as edges in the finite element mesh with sufficient damage. This fracture network covers a 6m$^2$ area evolving over the course of 7 milliseconds.  (b) Nodes represent fractures, with node size signifying total fracture length.  Edges represent fracture coalescence.} \label{hossfig}
\end{figure}

For each simulation, the transformation from finite element data to a fracture graph (see figure~\ref{hossfig}) reduces the data size from about 50GB to 15MB, and takes approximately 3 hours of processing time.  Simulations may be processed in parallel.

Given $N$ fractures, we encode the graph structure using a weighted adjacency matrix $A^{N \times N}$, where $A_{ij}$ represents the relationship between vertex $i$ and vertex $j$.  However, as different types of relationships between fractures can prove important, we use multiple distance metrics (say $d$ of them) to describe them, generalizing $A$ to a tensor of the form $A^{N\times N\times d}$.  Alternatively, we may interpret this as a node-aligned multilayer network \cite{kivela2014multilayer} with $d$ separate adjacency matrices $(A^{N\times N}_{(1)}, A^{N\times N}_{(2)}, \ldots, A^{N\times N}_{(d)})$.  Each of these defines a separate graph, but with shared vertex labels, allowing us to process each graph in parallel and then combine the results.  We also track a number of different attributes for each fracture, using the matrix $F^{N \times m}$.  The $d$ distance metrics and $m$ fracture attributes make up our feature set.

The input to the neural network will be a time series of $n$ graphs $(G_0,\ldots,G_{n-1})$, where at each time step $t$, $G_t$ consists of the $d$ adjacency matrices (of size $N \times N$) and the fracture attribute matrix (of size $N \times m$).  The output of the neural network will also be in exactly this format, but will consist of predictions; after having seen the first $t$ time steps of the graph, the neural network will predict time step $t+1$, and so on.  We thus train our network by comparing predictions to the true evolution of the fracture graphs.  Additionally, this allows a prediction for the evolution of a fracture graph, by iteratively predicting the next state given the starting state and the sequence of previous predictions.

Finally, we define the meaning of material failure in our graph formulation.  Failure refers to a system of fractures spanning the system from one boundary to the opposite boundary.  Since our fracture graph contains no explicit notion of boundary, we  introduce two new vertices representing the boundaries of the material.  We then say that the system has failed once there exists a fracture path (``failure path'') joining these two vertices.

\subsection{Graph Features}

Next, we describe the features of interest in our graph, associated both with \textbf{distances} between fractures and with \textbf{attributes} of fractures. We aim to predict the evolution of both categories of features over time, giving information both on individual fracture statistics and on overall damage to a material.

\subsubsection{Distance metrics}

We use $d=4$ distance metrics, each of which varies with time:

\begin{itemize}
\item{\textbf{Coalescence.}}
\label{ConnectivitySection}
This is a binary measure of connectivity.  A value of 1 represents coalescence of two fractures, while a value of 0 indicates that the fractures remain separate.  The quantity is symmetric: $A_{ij}=A_{ji}$ for fractures $i$ and $j$.

\item{\textbf{Minimal damage distance.}}
This is the smallest amount of additional damage needed to make two fractures coalesce.  Given damage threshold $\theta$, we assign a weight $\max[\theta-\text{damage},0]$ to each edge on the finite element mesh, and then find the minimum-weight path needed to link an element from one fracture to an element of the other.  This quantity is also symmetric.

\item{\textbf{Oriented tip distance.}}
We define a fracture tip as a finite element that belongs to the fracture but that is adjacent to only one other element in the fracture.  Based on the nearest tip at the previous time step, we approximate the fracture's direction of growth.  The oriented tip distance to another fracture is then the projected distance from this orientation to the first point of intersection with that fracture, or a large constant if no intersection occurs.  We include this metric on the assumption that fracture tip locations encode important information about the evolution of the fracture system.  Unlike the previous two quantities, it is asymmetric.

\item{\textbf{Shortest Euclidean distance.}}
We define the shortest distance between two fractures as the smallest possible Euclidean distance between an element in one and an element in the other.  Although this measure is not necessarily linked to whether two fractures will coalesce, previous studies~\cite{Tchoua2017} have noted that simple proximity of two fracture can be an important factor in fracture growth.  This quantity is symmetric.
\end{itemize}

Since the adjacency matrix represent similarities between fractures, for the last three metrics we adopt the standard procedure of using a Gaussian kernel to convert distances to similarities:
$$A_{ij}=e^{-\text{dist}^2_{ij}}.$$

\subsubsection{Fracture attributes} \label{featuresection}

Our feature set also includes attributes associated with each fracture, and that could be expected to influence fracture propagation.  They are represented by the matrix $F^{N \times m}$, where we use $m=13$ fracture attributes.  These attributes describe fracture length, damage, stresses experienced by fracture tips, fracture position (reflecting the observation that fractures often grow faster near the boundaries), orientation with respect to the loading direction, and connectivity.

\begin{itemize}
\item{\textbf{Total damage.}}  The sum of damage values on all edges associated with the fracture in the finite element mesh.
\item{\textbf{Projected \emph{x} length.}}  The projected Euclidean distance of the fracture along the $x$-axis, which is perpendicular to the loading direction of the material.
\item{\textbf{Projected \emph{y} length.}}  The projected Euclidean distance of the fracture along the $y$-axis, which is parallel to the loading direction of the material.
\item{\textbf{Euclidean length.}}  The largest Euclidean distance between any two tips of the fracture.
\item{\textbf{Total path length.}}  The sum of edge lengths associated with the fracture in the finite element mesh.
\item{\textbf{Maximum path length.}}  The largest sum of lengths of edges between any two tips of the fracture.
\item{\textbf{Max tip stress.}}  The largest stress at a tip of the fracture.  We define the stress at a tip as the average of the Cauchy normal stress values in the $y$-direction (loading direction), on all finite element edges touching the tip.
\item{\textbf{Mean tip stress.}}  The stress at a tip of the fracture, averaged over all its tips.
\item{\textbf{Mean fracture \emph{x} position.}}  The $x$-coordinate value of the center of mass of all edges associated with the fracture in the finite element mesh.
\item{\textbf{Mean fracture \emph{y} position.}}  The $y$-coordinate value of the center of mass of all edges associated with the fracture in the finite element mesh.
\item{\textbf{Fracture orientation.}}  The angle between the loading direction and the line that connects the tip with the smallest $x$-coordinate value to the tip with the largest $x$-coordinate value.
\item{\textbf{Fracture degree.}}  The number of coalescence events experienced by the fracture, equal to the sum of the elements in a row or column of the coalescence adjacency matrix.
\item{\textbf{Global failure.}}  A binary quantity, replicated as an attribute for every fracture in the system, indicating whether or not material failure has occurred.  This quantity reflects the arrest of fracture growth when stress is relieved by failure.
\end{itemize}

Note that the final attribute is used purely for training purposes, to help the neural network learn connectivity properties, and not for prediction.  Material failure is predicted based on the existence of a path of coalesced fractures connecting two boundaries.

\section{Machine Learning Methods}
\label{methodssection}

We use deep learning, based on large and diverse artificial neural networks~\cite{lecun2015deep}, to predict the growth of fractures.  Neural networks are a supervised learning method, requiring large amounts of training data.  We train on graph representations from multiple simulations of the evolving fracture system.  The goal is to predict, given representations of a single simulation at time steps $1, 2, \ldots, t$, the representation at time step $t+1$.  By iteratively applying this method, we can predict all future time steps as well.  We generally continue this process until the time step at which material failure occurs.

Deep learning has been applied successfully in the past to graph classification problems \cite{kipf2016semi} and to dynamic graph prediction problems \cite{manessi2017dynamic}. Many different forms of neural networks exist, and these different forms are frequently used in combination.  Convolutional neural network (CNN) models have been used successfully on graphs by using the graph Fourier transform to create a graph convolutional network (gCN)~\cite{DBLP:journals/corr/HenaffBL15}. Kipf and Welling have provided fast and accurate methods for convolutions on large graphs \cite{kipf2016semi}, using approximations to the eigen decomposition of the graph Laplacian.  Manessi has proposed two promising methods of combining gCNs with recurrent neural networks (RNNs), to exploit graph data with temporal information~\cite{manessi2017dynamic}.  The neural network architecture that we use here is made up of several components including feed-forward networks, graph convolutional networks (gCNs), and recurrent neural networks (RNNs).

\subsection{Feed-Forward Neural Networks} \label{fnnsection}

The basic unit in a neural network is an artificial neuron that receives inputs from a fixed number of sources and decides its output based on a weighted average of its inputs.  These neurons are combined as layers, the simplest being a \textit{feed-forward layer} or a \textit{fully-connected layer}, where neurons are independent and process all of the input simultaneously.  The resulting output then is primarily a series of matrix multiplications.  Given a vector input $X^{m\times 1}$ to the network, the first layer's vector output $Y^{k\times 1}$ may be expressed as
$$Y = \sigma(WX + B).$$
Here, $\sigma$ is a simple (though generally nonlinear) activation function such as the rectified linear unit, or ReLU, given by $\sigma(x) = \max{(0,x)}$.  $W$ is an $k\times m$ weight matrix, and $B$ is a vector bias allowing the neuron to be activated at an adjustable threshold value.  Neural network architectures used in deep learning can consist of many such ``hidden'' layers, with a feed-forward network representing a composition of activation functions.  The key observation is that adding additional layers (depth) to a neural network can allow it to approximate a function far more effectively than adding additional neurons to a given layer~\cite{hornik1989multilayer,lin2017does}.

The main challenge of a neural network lies in training its adjustable parameters.  In the training phase, labeled data are repeatedly passed through the network to determine the error of a given prediction (``loss function'').  The weights $W$ and biases $B$ are then adjusted in order to minimize the loss in the training set.  Using backpropagation to attribute a specific part of the loss to each neuron, a gradient is calculated, allowing the prediction to be optimized iteratively through a form of stochastic gradient descent.

\subsection{Graph Convolutional Networks (gCN)} \label{gcnsection}
A recently developed neural network architecture particularly well suited to operations on graphs~\cite{kipf2016semi,DefferrardBV16,manessi2017dynamic} involves using a convolutional layer.  The goal of a convolution is to provide spatial invariance: in image processing, for instance, convolving an image with a small kernel can enable recognition of image features regardless of where they are located within the image~\cite{krizhevsky2012imagenet}.  Similar methods can be used for graph data.  Recall that the input to our neural network is a series of graphs, each defined by an adjacency matrix $A^{N\times N}$ and a node feature (attribute) matrix $F^{N\times m}$.  A graph convolutional layer~\cite{kipf2016semi} with input $F$ results in the $N \times k$ output
$$Y = \sigma\left(\tilde{D}^{-\frac{1}{2}}\tilde{A}\tilde{D}^{-\frac{1}{2}}FW\right), $$
where $\tilde{A} = A + I $ is the adjacency matrix with added self-connections, $\tilde{D}$ is a diagonal matrix with $\tilde{D}_{ii} = \sum_{j} \tilde{A}_{ij}$, and $W$ is an $m\times k$ weight matrix.  $\tilde{D}^{-\frac{1}{2}}\tilde{A}\tilde{D}^{-\frac{1}{2}}$ is the symmetric normalized graph Laplacian, which takes an average of surrounding vertices and approximates a convolution with the adjacency matrix, so that the graph structure itself forms a filter for the weight matrix. The output $Y$ from the gCN layer consists of $k$ newly computed attributes associated with each of the $N$ nodes in the graph.  The use of multiple layers, each with its own weight matrix, allows the use of information from neighborhoods consisting of multiple adjacency steps in the graph.

\subsection{Recurrent Neural Networks (RNN)} \label{rnnsection}

RNNs are structurally similar to feed-forward networks, but rather than simply mapping an input $X$ to an output $Y$, they map a sequence of inputs $X_1, X_2, \dots, X_n$ to a sequence of outputs $Y_1, Y_2, \dots, Y_n$.  Furthermore, they carry forward an internal state, allowing them to learn long-term dependencies between events if the input sequence represents a time series.  At time step $t$, they take the state produced by the previous step, $S_{t - 1}$, together with $X_t$ to produce a new state, $S_t$, and an output $Y_t$.  The simplest form of RNN uses an activation function
$$ S_t = \sigma(W_S S_{t-1} + W_X X_t) $$
and
$$ Y_t = \sigma(W_Y S_t), $$
where $W_S$, $W_X$, and $W_Y$ are different weight matrices that are adjusted during training. The distinction between a feed-forward network and a RNN may be understood graphically as the network ``unrolling'' across time (Figure \ref{RNNArchitecture}).  The internal state $S_t$ represents the RNN's ``memory'' of the system, constructed from the previous memory together with the new information entering the network.

\tikzset{%
  >={Latex[width=2mm,length=2mm]},
            base/.style = {rectangle, rounded corners, draw=black,
                           minimum width=1cm, minimum height=1cm,
                           text centered, font=\sffamily},
  layer/.style = {base, fill=green!30},
  input/.style = {base, fill=blue!30},
  output/.style = {base, fill=red!30},
  textbox/.style = {minimum width=1cm, minimum height=1cm,
                           text centered, font=\sffamily},
}

\begin{figure}[!htb]
\begin{center}
\captionsetup[subfigure]{labelformat=empty,justification=raggedright,singlelinecheck=false}
\begin{subfigure}{0.29\textwidth}
\scalebox{0.70}{\hspace{40pt}
\begin{tikzpicture}[node distance=1.5cm,
    every node/.style={fill=white, font=\sffamily}, align=center]
  \node (ff)             [layer]              {Layer};
  \node (ffinput)     [input, below of=ff]          {$X$};
  \node (ffoutput)      [output, above of=ff]   {$Y$};  
  \draw[->]           (ffinput) -- (ff);
  \draw[->]                (ff) -- (ffoutput);
  \end{tikzpicture}
 }
\caption{\hspace{12pt}Feed-forward}
\end{subfigure}
\captionsetup[subfigure]{labelformat=empty,justification=centering,singlelinecheck=false}
\begin{subfigure}{0.7\textwidth}

\scalebox{.70}{
\begin{tikzpicture}[node distance=1.5cm,
    every node/.style={fill=white, font=\sffamily}, align=center]
  \node (rec1)             [layer]              {Layer};
  \node (recinput1)     [input, below of=rec1]          {$X_1$};
  \node (recoutput1)      [output, above of=rec1]   {$Y_1$};
  \node (rec2)             [layer, right = 1.5cm of rec1]              {Layer};
  \node (recinput2)     [input, below of=rec2]          {$X_2$};
  \node (recoutput2)      [output, above of=rec2]   {$Y_2$};  
  \node (rec3)             [layer,right = 1.5cm of rec2]              {Layer};
  \node (recinput3)     [input, below of=rec3]          {$X_3$};
  \node (recoutput3)      [output, above of=rec3]   {$Y_3$};  
  \node (recdot)             [textbox, right = 1.5cm of rec3]              {\ldots};
  \node (recn)             [layer, right = 1.5cm of recdot]              {Layer};
  \node (recinputn)     [input, below of=recn]          {$X_t$};
  \node (recoutputn)      [output, above of=recn]   {$Y_t$};  
  \draw[->]           (recinput1) -- (rec1);
  \draw[->]                (rec1) -- (recoutput1);
  \draw[->]           (recinput2) -- (rec2);
  \draw[->]                (rec2) -- (recoutput2);
  \draw[->]           (recinput3) -- (rec3);
  \draw[->]                (rec3) -- (recoutput3);
  \draw[->]           (recinputn) -- (recn);
  \draw[->]                (recn) -- (recoutputn);
  \draw[->]			       (rec1) -- node {$S_1$} (rec2);
  \draw[->]			       (rec2) -- node {$S_2$} (rec3);
  \draw[->]			       (rec3) -- node {$S_3$} (recdot);
  \draw[->]			       (recdot) -- node {$S_{t-1}$} (recn);
  \end{tikzpicture}
}
\caption{\hspace{-15pt}Recurrent}
\end{subfigure}
\end{center}
\caption{Feed-forward vs recurrent neural networks}\label{RNNArchitecture}
\end{figure}
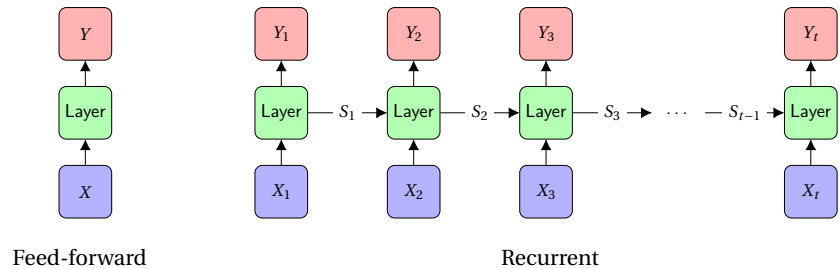

In practice, the exact operations used to compute $S_t$ and $Y_t$ depend on the RNN implementation.  We use the most common method, known as long short-term memory (LSTM)~\cite{hochreiter1997long}.  This includes additional weights allowing states to be ``forgotten'' while also explicitly using the previous time step's output $Y_{t-1}$ in the calculation at step $t$.  Importantly, RNNs can be incorporated as layers in networks that also includes feed-forward and gCNs, and can be trained using the same methods as these.

\subsection{Network Architecture}\label{architecturesection}

Our architecture combines the components discussed above.  The input is a time series of graphs $G_0,\dots, G_{n-1}$, each consisting of an adjacency matrix $A$ and a feature matrix $F$.  The output is a series of the same length, but shifted over one time step: $G_1,\dots, G_n$, with the network predicting a given $G_t$ based only on the information up to $G_{t-1}$.  This allows us to train the network by comparing the predictions for $G_1,\dots,G_n$ with the actual sequence graphs.

\tikzset{%
  >={Latex[width=2mm,length=2mm]},
            base/.style = {rectangle, rounded corners, draw=black,
                           minimum width=1cm, minimum height=1cm,
                           text centered, font=\sffamily},
  layer/.style = {base, fill=green!30},
  rnnout/.style = {base, fill=yellow!30},
  input/.style = {base, fill=blue!30},
  output/.style = {base, fill=red!30},
  textbox/.style = {minimum width=1cm, minimum height=1cm,
                           text centered, font=\fontsize{11}{13.2}\sffamily,},
}
\begin{figure}[!b]
\centering
\scalebox{0.7}{
\begin{tikzpicture}[node distance=2cm,
    every node/.style={fill=white, font=\sffamily}, align=center]
\node [input]                        (A)           {$F_t, A_t$}; 
\node [layer, above of = A]            (GC)          {gCN};
\node [layer, above of = GC]           (RNN)         {RNN};
\draw[->]			  	  (-4,4) -- node {$S_{t-1}, Y_{t-1}$} (RNN);
\draw[->]			      (RNN) -- node {$S_{t}, Y_{t}$} (4,4)  ;
\node (dots1)             [textbox, left = 3cm of RNN]              {\ldots};
\node (dots2)             [textbox, right = 3.5cm of RNN]              {\ldots};
\node [rnnout, above left of = RNN]   (Yt)          {$Y_t$};
\node [layer, above of = Yt]          (FF)          {FF};
\node [output, above of = FF]         (F_plusone)   {$F_{t+1}$};
\node [rnnout, above right of = RNN]  (cross)       {$A_t, Y_t$};
\node [layer, above of = cross]        (conv)        {$1\times1$ Conv};
\node [output, above of = conv]       (A_plusone)   {$A_{t+1}$};

\draw[->] (A.east) to [out=345, in=270] (cross);

\draw [arrows=-Stealth] 
     (A)           edge (GC)
     (GC)          edge (RNN)
     (RNN)         edge (Yt)
     (Yt)          edge (FF)
     (FF)          edge (F_plusone)
     (RNN)         edge (cross)
     (cross)       edge (conv)
     (conv)        edge (A_plusone)
;
\end{tikzpicture}
}

\caption{Outline of one "unrolled" computation section of the algorithm, showing four different neural network processing units: a gCN, an RNN, a feed-forward network, and a $1 \times 1$ convolution.}
\label{fig:arch_outline}
\end{figure}
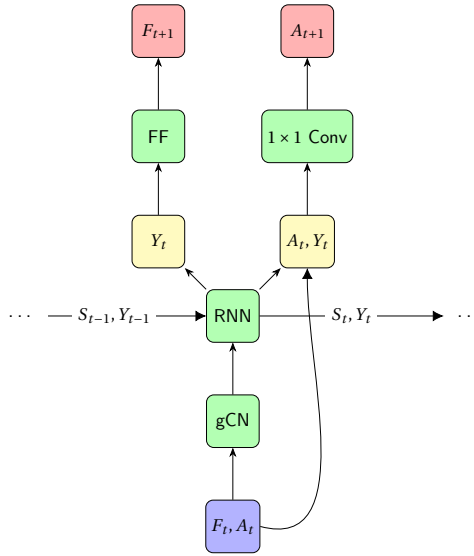

Figure \ref{fig:arch_outline} summarizes the procedure for a given time step $t$.  $F_t$ represents the (input) attribute matrix based on fractures, and $A_t$ represents the (input) adjacency matrix based on distance metrics.  Following the methods in Manessi et al.~\cite{manessi2017dynamic}, the gCN extracts relevant features and the RNN creates a ``memory'' stored in $S_t$.  These two layers together compute, for each vertex, a set of new features described by the output $Y_t$, taking into account prior developments in the graph.  Since we have $d$ different distance metrics, $A_t$ is actually a tensor of dimensions $N \times N \times d$.  This is inconsistent with the definition of a graph convolutional layer from \cite{kipf2016semi}, which assumes a single $N \times N$ adjacency matrix.  We therefore conduct $d$ independent graph convolution operations in parallel and combine the results, ultimately obtaining a concatenated output matrix $Y_t$ of size $N \times kd$, where $k$ is an adjustable parameter corresponding to the number of output features in each graph convolution.
We pass $Y_t$ directly into a feed-forward layer to predict the fracture attribute values $F_{t+1}$ at the next time step.

However, predicting the evolution of the adjacency matrix is not straightforward using existing approaches, since gCNs use graph structure to predict vertex features rather than to predict edge features.  We therefore augment $A_t$ with the RNN output $Y_t$.  Since the adjacency matrix has dimensions $N \times N \times d$, each element associated with fractures $i$ and $j$ can be considered as a $d$-dimensional vector.  We append to this vector the output features associated with both fractures, creating a new vector of length $d+2kd$.
%

We would like to process each of these vectors independently of the others.  In order to do so, we apply a $1 \times 1$ convolution to the augmented ($N \times N \times d+2kd$) tensor.  This is roughly equivalent to applying a feed-forward layer to each vector, changing its length from $d+2kd$ to a length of our choosing.  By setting the length to $d$, we recover the correct dimensions for the adjacency matrices.  Analogously to the case of fracture attributes, we then use the $1\times 1$ CNN to predict $A_{t+1}$ at the next time step.

Each of the layers in our architecture requires training a set of weights and biases.  Consider an adjustable weight $w$, and a loss function that is differentiable with respect to $w$.  (Note that even if we are trying to predict a discrete quantity such as binary coalescence, the output of the neural network will be a real number that we ultimately threshold, and so an appropriately chosen error will be continuous and differentiable.)  There exist both momentum-based algorithms, such as gradient descent, and norm-based algorithms, such as RMSProp, that allow us to update $w$ such that its new value will reduce the loss. We use Nesterov Accelerated Adaptive Moment (Nadam) Estimation~\cite{Ruder2017}, which combines standard momentum (using a decaying mean instead of a decaying sum) with RMSProp to improve performance.

We implement the architecture above in Python, using Keras for all components except for the gCN, for which we use the implementation from~\cite{kipf2016semi}.  In each component, we use two hidden layers, with 64 neurons per layer.  Additionally, our use of graphics processing units (GPUs) accelerates training by up to two orders of magnitude
as neural network computations, like graphical rendering, consist mostly of large matrix multiplications.

\section{Results and Discussion}
\label{resultssection}

\subsection{Training}

Our training data consist of 145 HOSS simulations, as described in Section~\ref{physicalsystemsection}.
The simulations include time steps 0 through 350 representing the 7 milliseconds of fracture evolution, though some simulations end earlier, right after the material fails.  In those cases, we pad the missing time steps by repeating the final state, since fracture systems rarely exhibit any significant change after stress is relieved through material failure.

We train using 5-fold cross-validation. This framework involves partitioning the 145 simulations into 5 subsets, or {\em folds\/}, of equal size.  We then train 5 separate networks, each one using the data from one of these folds as the test set (29 simulations) and the remaining data as either the training set (101 simulations) or validation set (15 simulations).  In this way, collectively over the 5 networks, we ultimately use all of our simulations as test data.  For each network, we initialize weights
as uniform random variables, and then iteratively update them by training on the 101 simulations, in random order.  In each simulation,
we sequentially cycle through time steps $t=0,\dots,349$, inputting graph $G_t$ at each time step.  The output is a predicted time series $(G^{\mathrm{out}}_1,\dots,G^{\mathrm{out}}_{350})$ which we compare with the actual graphs.  From this comparison, we calculate a loss value as the sum of the mean squared error (MSE) of all features, and use backpropagation to adjust the weights in the network.  All real-valued features are rescaled (in preprocessing) to have mean zero and variance one, so they carry equal weight in the loss function.  A single training pass on all simulations is called one {\em epoch\/}.  With every new epoch, a new random permutation of the simulations is generated.

Fracture propagation in HOSS simulations is typically sporadic, with rapid evolution followed by long periods of little or no change.  These discontinuities are challenging for a neural network, leading it to overfit by learning only the dormant behavior or learning propagation patterns so abrupt as to be unphysical. To overcome problems such as these, we use three different forms of regularization:

\begin{enumerate}
\item {\bf Downsampling.}
We use only every $25$th time step for both training and prediction, relabeling time steps $0, 25, \dots, 350$ as $0, 1, \dots, 14$.  This has the effect of making fracture evolution more apparent to the network.

\item {\bf Early stopping.}
After every 5 epochs, which we call one ``iteration,'' we calculate the training error on a separate validation set.  If this error starts to increase rather than decrease (see Fig.~\ref{fig:learning}), that signifies possible overfitting to the training set, and so we halt training.  Otherwise, we continue training for 40 iterations.

\begin{figure}[h]
	\centering
	\includegraphics[width = 0.8\textwidth]{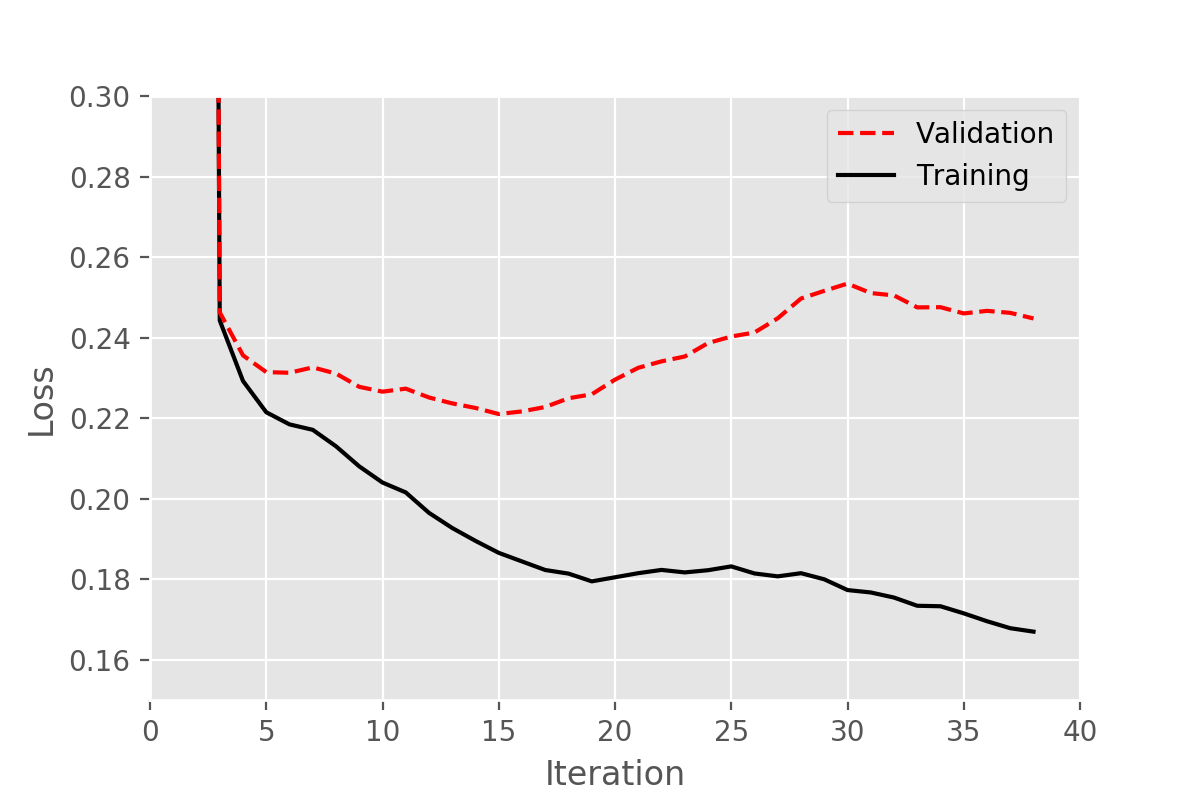} 
	
	\caption{Learning curve for one network, showing evolution of loss values on training and validation sets (smoothed with moving average using window of width 5 iterations).  While training loss is almost always decreasing, validation loss shows marked increase after about 15 iterations, indicating possible overfitting.} \label{fig:learning}
\end{figure}

\item {\bf Data augmentation.}
The size of our training data is modest by deep learning standards, and further reduced by downsampling. To compensate, we develop an innovative use of synthetic data.  After the first iteration, we allow the network to train on its own latest predictions for (downsampled) time steps 1 through 14 of each simulation.  We do this by appending those predictions to our real data during the second iteration: we expand the feature set from $d$ distance metrics to $2d$ distance metrics ($d$ of them real, $d$ of them synthetic) and from $m$ attributes to $2m$ attributes ($m$ of them real, $m$ of them synthetic).  With each successive iteration, we append an additional set of the most recently generated predictions: after $k$ iterations, the feature set will include $(k+1)d$ and $(k+1)m$ features, all of which we use in training.

With the process repeated in this way, the network learns to correct patterns of incorrect prediction, since the loss is always calculated by comparing the output with actual simulation data.  In order to decrease the importance of the less accurate earlier predictions, we assign a weight to the loss function contributions that is inversely proportional to the age of the synthetic data: the latest ($k$th) set has weight 1, the previous set has weight $1/2$, and so on back to the oldest set with weight $1/k$.  The improvement in prediction quality due to this method is seen in Fig.~\ref{fig:SynthDataFig}, showing predicted total fracture damage in one reference simulation (discussed further below).  With synthetic data, quantitative predictions track the actual fracture damage far more closely, in terms of both mean and standard deviation.

\begin{figure}
	\centering
	\includegraphics[width = 0.49\textwidth]{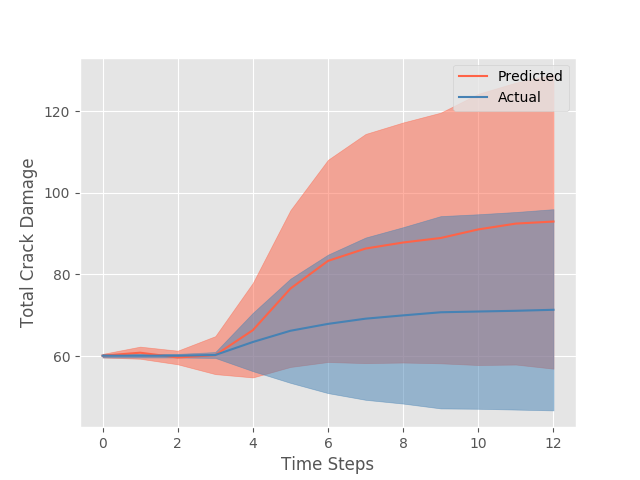} 
	\includegraphics[width = 0.49\textwidth]{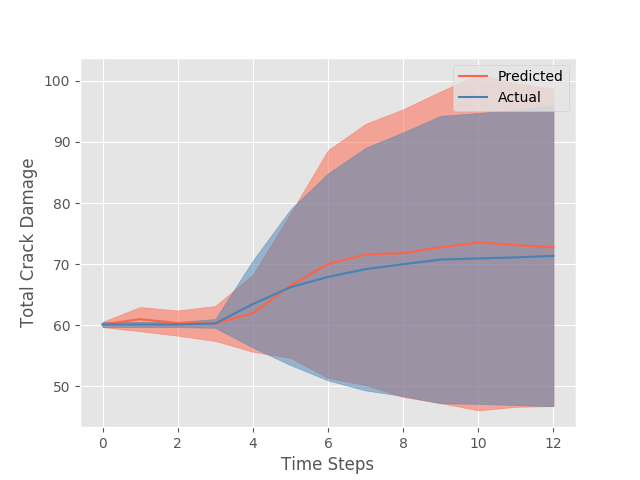}\\
    
    (a) \hspace{2.15in} (b)
    
	\caption{Evolution of predicted fracture damage vs.\ actual (i.e.,\ simulated) fracture damage, in one reference simulation,
without (a) and with (b) the use of data augmentation.  The shaded area indicates one standard deviation from the mean, measured across all fractures in the system.} \label{fig:SynthDataFig}
\end{figure}
\end{enumerate}

The entire training procedure, as described above, takes approximately 6 hours of processing time on a GPU.

\subsection{Prediction}

We generate predictions of the entire time series of fracture evolution, based on $t=0$ conditions alone, for all 145 simulations (each of which is in the test set of one of our 5 trained networks).  These predictions are of two kinds: the evolution of fracture sizes, and the global measure of when material failure occurs.  On a trained neural network, we typically obtain predictions within seconds.

\subsubsection{Fracture Growth Statistics}

\begin{figure}
	\centering
	\includegraphics[width = \textwidth]{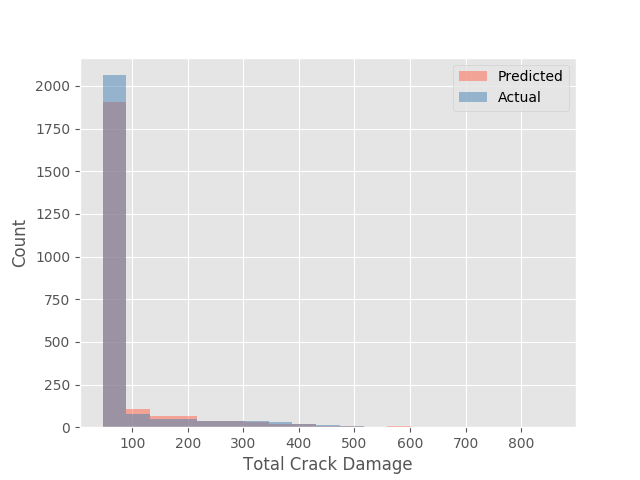}\\
	\caption{Distribution of total damage in coalesced fractures (predicted and actual), aggregated over all simulations.}  \label{TotalDamageHistFig}	
\end{figure}

One measure of fracture size is the total amount of damage associated with the fracture.  Fig.~\ref{fig:SynthDataFig} above describes the evolution of predicted and actual (meaning simulated) fracture damage.  The mean and standard deviation are taken over all fractures in one reference simulation, with the large standard deviation in part reflecting the concentration of damage among a limited number of fractures.  However, recall that there is arbitrariness in the method used to identify individual fractures once they have coalesced (see Section~\ref{fracturegraphsection}).  Therefore, we also consider damage in connected components of coalesced fractures, focusing specifically on the final time step of a simulation.  Fig.~\ref{TotalDamageHistFig} shows the distribution of total damage in these coalesced fractures, aggregated over the entire set of 145 simulations.  The predicted and actual histograms are both dominated by the peak on the left that represents the overwhelming majority of fractures with only minimal damage.  While there is some slight underprediction of coalesced fractures with the least damage, the predicted mean is within 2.45\% of its simulated value.

\begin{figure}
	\centering
	\includegraphics[width = 0.49\textwidth]{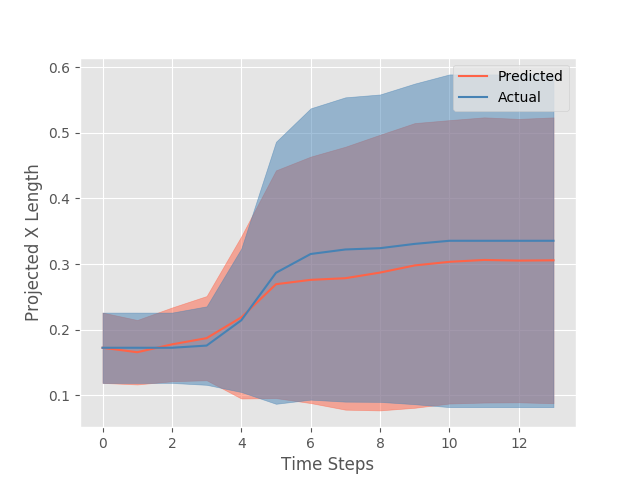}
	\includegraphics[width = 0.49\textwidth]{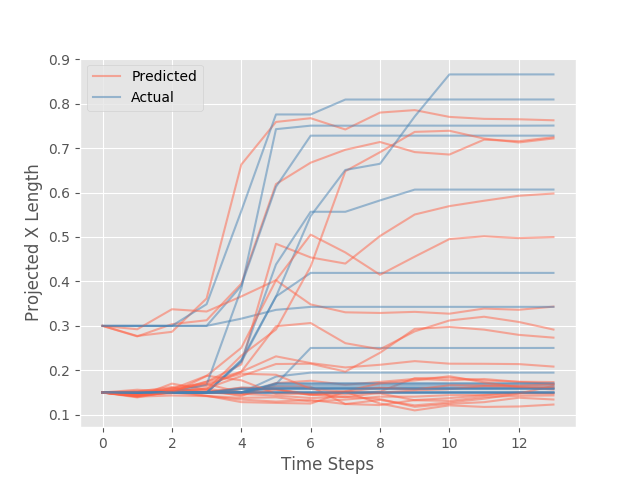}\\
	    
    (a) \hspace{2.15in} (b)
    
	\caption{Evolution of projected $x$ length of a fracture (predicted and actual) for the simulation from Fig.~\ref{hossfig}.  In (a), the shaded area indicates one standard deviation from the mean, measured across all fractures in the system.  In (b), each individual fracture is plotted.}  \label{Lengthfigure}
\end{figure}

Other measures of fracture size include its length attributes.  Fig.~\ref{Lengthfigure} shows the predicted evolution of projected $x$ length (perpendicular to the loading direction) of fractures, in the reference simulation used so far.  Here, the standard deviation is significant even at $t=0$, since the initial distribution is bimodal, reflecting whether a fracture is oriented horizontally or diagonally (see Fig.~\ref{hossfig}).  Our predictions, while not perfect, agree with many aspects of the actual fracture length evolution, both in terms of overall statistics and in terms of individual fracture behavior.

\begin{figure}
	\centering
	\includegraphics[width = \textwidth]{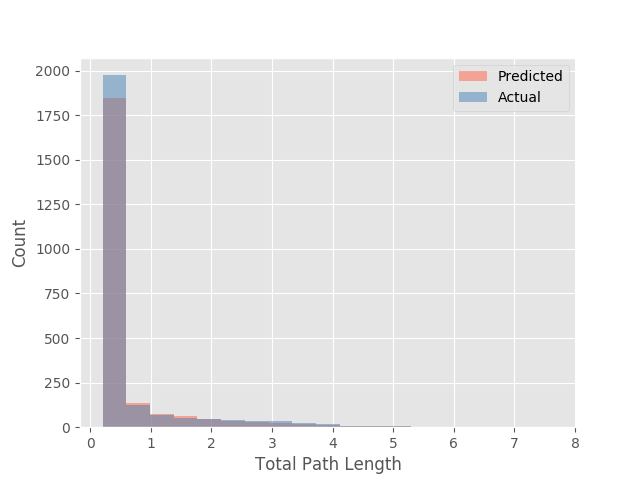}\\
	\caption{Distribution of total path length in coalesced fractures (predicted and actual), aggregated over all simulations.}  \label{TotalPathLengthHistFig}	
\end{figure}

\begin{figure}
	\centering
	\includegraphics[width = 0.49\textwidth]{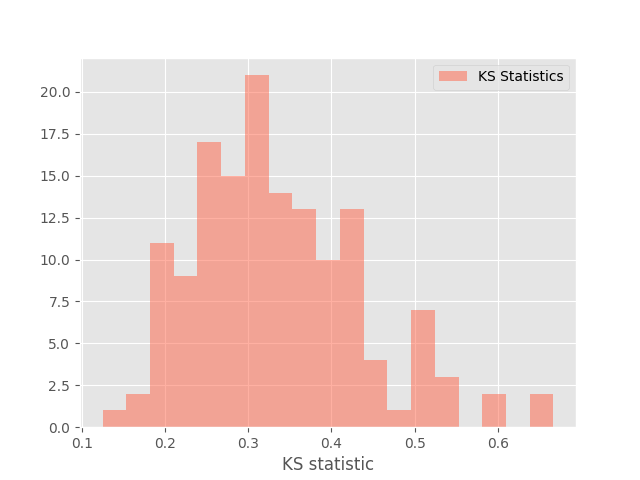}
	\includegraphics[width = 0.49\textwidth]{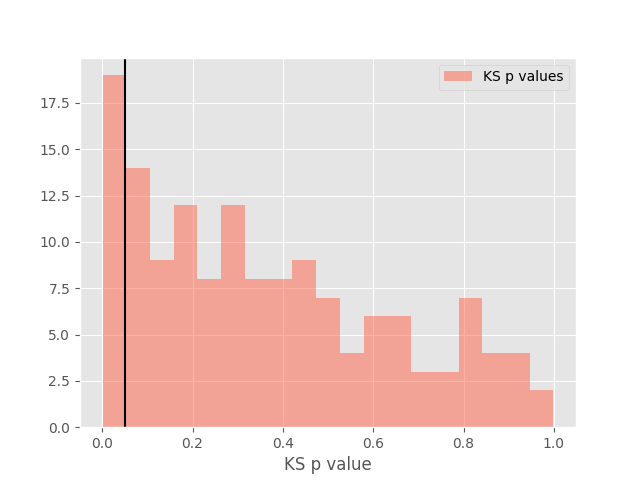}\\
	    
    (a) \hspace{2.15in} (b)

	\caption{Histogram, over all simulations, of (a) the KS statistic measuring the difference between predicted and actual distributions of total path lengths in coalesced fractures, and (b) the $p$-value corresponding to the KS statistic.  Vertical line shows value above which distribution difference is not significant at a level of $\alpha=0.05$.}
    \label{KSpvals}	
\end{figure}

A length attribute that is directly relevant to material failure but also very closely related to a coalesced fracture's total damage is its total path length. Fig.~\ref{TotalPathLengthHistFig} shows the distribution of this quantity at the final time step, aggregated over the entire set of simulations.  Not surprisingly, it strongly resembles Fig.~\ref{TotalDamageHistFig}, with the predicted and actual histograms again displaying a similar initial peak representing unpropagated fractures.  The predicted mean is 1.85\% above its simulated value.  The difference between the two histograms is described quantitatively by the Kolmogorov-Smirnov (KS) statistic in Fig.~\ref{KSpvals}: KS distances are concentrated at smaller values, with a $p$-value that lies above 0.05 for 126 out of 145 simulations.  Thus, in most cases, the predicted total path length distribution is consistent with the actual one.
Surprisingly, this result is robust under ablation, or exclusion of individual features or feature groups during training~\cite{nguyen2015deep}.  We find that the \emph{only} quantity whose removal increases the mean KS distance significantly (from 0.328 to 0.374) is fracture position.  This presumably reflects the fact that position can be a determining factor for coalescence, based on whether the fracture is nearer to the boundary or to the center of the material.  Conversely, our other features appear to contain enough redundancy that the network can easily make up for the absence of some of them.

%

\subsubsection{Material Failure}

One of the main strengths of our deep learning approach is that it can provide predictions of global effects such as material failure at the same time as individual fracture statistics.  We determine whether a material has failed by whether there exists a connected component of coalesced fractures that includes both boundary fractures.  However, a straightforward application of our trained network is not sufficient to obtain accurate results: parameter settings that give accurate predictions of fracture statistics can often predict that the material fails far later than it should. 

We therefore modify certain parameters, and also apply a postprocessing transformation.
The crucial parameters involve a coalescence threshold: since the predicted coalescence adjacency matrix $A^{\textrm{out}}$ is real-valued, we consider that fractures $i$ and $j$ have coalesced if either $A^{\textrm{out}}_{ij}$ or $A^{\textrm{out}}_{ji}$ exceed a fixed threshold value.  In our previous results, the coalescence threshold for synthetic data used in the data augmentation step of training was set by default to 0.5, and the threshold used in prediction was set to 0.1.  Raising the training threshold makes coalescence prediction more aggressive, as it pushes the network to correct its perceived underprediction in the data augmentation process.  By contrast, raising the prediction threshold makes coalescence prediction more conservative.  We find that raising both thresholds simultaneously provides the best results: we set the training threshold to 0.9 and the prediction threshold to 0.5.

Of the original 145 simulations on which we make predictions in our 5 trained networks, 84 of them do not exhibit failure by the final time step.  In those cases, actual (simulated) time to failure is undefined, so we omit them from our analysis and only obtain predictions for the remaining 61 simulations.  When predicting time to failure, we allow the network to run beyond time step 14 if necessary.  We then perform an additional postprocessing step using ridge regression (Tikhonov regularization)
and leave-one-out cross-validation, where for each one of the 61 simulations, we learn a separate transformation, based on the actual times to failure and the network predictions for the 60 other simulations.  In principle, this method could entail a small amount of data leakage: the transformation is learned from results from all 5 trained networks, 4 of which were trained on data that included the test simulation itself.  However, given that a network is trained on a set of 101 simulations, it appears unlikely that one single simulation could influence the results enough to bias the transformation measurably.


\begin{figure}[t]
	\centering
	\includegraphics[width = \textwidth]{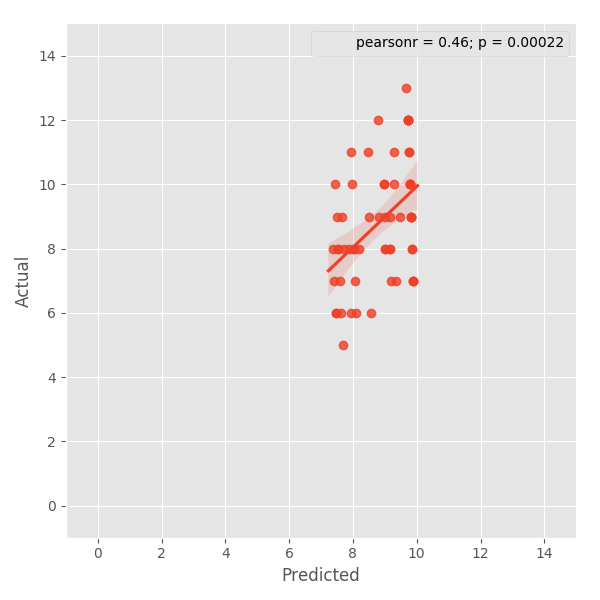}
	    

	\caption{Scatter plot of predicted versus actual time to failure, for the 61 fracture systems that exhibit material failure during the simulation.
	MAE is 15.4\%.
	}  
    \label{scatterPred}	
\end{figure}

Fig.~\ref{scatterPred} shows a scatter plot with predicted vs.\ actual time to failure for the 61 fracture systems, after applying the postprocessing transformation.  The mean absolute error (MAE) in the prediction is 15.4\% of the average time to failure, and the correlation coefficient is $r=0.46$.  This result is competitive with the most recent ML approaches~\cite{Moore2018} based on similar HOSS data, which have an error of approximately 16\% and a correlation coefficient varying from $r=0.42$ to $r=0.68$, even though those methods (unlike ours) are specifically oriented towards predicting failure paths.  The MAE in our results reflects the non-negligible scatter in predicted values for a given actual time to failure, which may arise in part from our features imperfectly characterizing the geometry of fractures.  While this issue would not impact predictions of fracture growth statistics, which directly measure the feature values themselves, it could easily cause imprecision in predicting higher-order effects such as material failure.  Our postprocessing will correct any systematic overestimate or underestimate in predicted time to failure (see regression line in Fig.~\ref{scatterPred}), but it will not reduce the variability in prediction.  It is worth noting, however, that this phenomenon is hardly unique to our approach. Time to failure is notoriously difficult to predict even with high-fidelity models: the quantity exhibits large sample-to-sample variations in the original HOSS simulation data~\cite{Rougier2014}, even when other fracture growth statistics are relatively stable.


\section{Conclusions}
\label{conclusions}

We have presented a novel ML approach that provides rapid and accurate predictions of fracture propagation in brittle materials. Using deep learning, we construct a neural network architecture that combines a gCN with an RNN, and train the network on high-fidelity simulations from the HOSS model. We compensate for the relatively modest size of our training data set with a new form of data augmentation, training the network not only on simulation results but also on incorrect predictions that it has itself generated in earlier training passes. In this way, the network learns to correct its own mistakes, and successfully predicts how properties of the fracture system evolve from their $t=0$ state alone.  A significant benefit of an ML approach is speed: whereas a single HOSS simulation takes hours to run, our neural network generates results in seconds once it has been trained. This speed consideration is particularly crucial in the context of uncertainty quantification.  HOSS simulations themselves represent random realizations of a statistical ensemble, and thousands of such simulations could be required to produce accurate estimates of material behavior.

Our fracture size predictions, which we test using 5-fold cross-validation, are in good agreement with the results of simulations.  The predicted total damage in a coalesced fracture at the final time step, averaged over all 145 simulations, is within 2.95\% of its simulated value.  The predicted total length of a coalesced fracture is within 1.85\% of its  simulated value.  In 87\% of the simulations, the difference between the predicted and simulated length distribution is not statistically significant, at the $0.05$ significance level of a Kolmogorov-Smirnov test.
While other ML approaches have also proven successful for modeling fracture dynamics, deep learning has the advantage that it can generate predictions simultaneously for qualitatively distinct material properties.  In particular, we predict time to material failure with an accuracy that is competitive with the state-of-the-art.  The MAE in our predictions is approximately 15\% of the average time to failure, with a correlation coefficient of $0.46$ between simulated times and predicted times.


It is important to note that the results presented in this work are based on high-fidelity models that have been run under a given type of boundary and initial conditions. If different conditions were to be addressed, then the process would need to be repeated. Ideally, one would expect to devise a matrix or a space of boundary and initial conditions that would cover the most relevant situations of interest for a particular problem. This would then create a library of high-fidelity simulation results used to inform machine learning algorithms, for predicting more general loading scenarios.

Finally, we mention possible directions for improving the quality of results.  Predicted fracture sizes are stable under exclusion of different groups of fracture attributes, suggesting redundancy in the current choice of features. The use of new features, representing different physical properties, may improve predictions.  Additionally, time-to-failure predictions could possibly be improved by the use of ensemble learning, combining the results of multiple trained networks with differently adjusted parameters.  This is due to a significant class imbalance when predicting the coalescence of individual fractures: only a small minority of fracture pairs ever coalesce.
Preliminary studies suggest that these pairs may be more accurately identified by using an ensemble of networks trained with different coalescence threshold settings, with an additional supervised learning step to map their multiple predictions onto a single one.
Alternatively, a reweighting scheme could be used in such cases~\cite{Moore2018}, so that a classifier has adequate opportunities to learn cases of coalescing fractures.  Implementing class reweighting could potentially improve coalescence predictions to the point where a single model would provide highly accurate results for time to failure.


\section{Acknowledgments}
The authors are grateful to an anonymous reviewer, whose comments and suggestions helped improve the clarity of the manuscript.
This work was supported by the U.S. Department of Energy at Los Alamos National Laboratory under Contract No.~DE-AC52-06NA25396 through the Laboratory Directed Research and Development program (award 20170103DR).

\section*{Data Availability}

The raw/processed data required to reproduce these findings cannot be shared at this time due to technical or time limitations.

\section*{References}

\bibliography{mybibfile}

\end{document}